\documentclass{aa}  

\usepackage{graphicx}
\usepackage{txfonts}
\usepackage{lipsum}
\usepackage{subcaption}         % necessary for continued figures, example in section 3
\usepackage{lscape}             % to rotate a single page table, example in appendix.
\usepackage{placeins}           % useful with \FloatBarrier, to keep 
\usepackage{ulem}
\usepackage{bm}
\usepackage{color}
\usepackage{amssymb}
\usepackage{amsmath}
\usepackage{graphicx}
\usepackage{amsfonts}
\usepackage{float}
\usepackage{array}
\allowdisplaybreaks
\usepackage{dcolumn}
\usepackage{epsf}
\usepackage{tabularx}
\usepackage{soul}
\usepackage{wrapfig}
\usepackage[colorlinks=true, linkcolor=red, citecolor=blue, filecolor=cyan, urlcolor=magenta]{hyperref}
\usepackage{multirow}

\usepackage{color}
\usepackage{natbib,twoopt}
\usepackage[hyphenbreaks]{breakurl}
\bibpunct{(}{)}{;}{a}{}{,}             %% natbib format for A&A and ApJ
\definecolor{cobalt}{rgb}{0.06, 0.2, 0.65}
\hypersetup{
  colorlinks,
  citecolor=cobalt,
  linkcolor=[rgb]{0.8, 0.2, 1.0},
  urlcolor=cobalt,
}
\makeatletter
  \newcommandtwoopt{\citeads}[3][][]{\href{http://ui.adsabs.harvard.edu/abs/#3}%
    {\def\hyper@linkstart##1##2{}%
     \let\hyper@linkend\@empty\citealp[#1][#2]{#3}}}
  \newcommandtwoopt{\citepads}[3][][]{\href{http://ui.adsabs.harvard.edu/abs/#3}%
    {\def\hyper@linkstart##1##2{}%
     \let\hyper@linkend\@empty\citep[#1][#2]{#3}}}
  \newcommandtwoopt{\citetads}[3][][]{\href{http://ui.adsabs.harvard.edu/abs/#3}%
    {\def\hyper@linkstart##1##2{}%
     \let\hyper@linkend\@empty\citet[#1][#2]{#3}}}
  \newcommandtwoopt{\citeyearads}[3][][]%
    {\href{http://adsabs.harvard.edu/abs/#3}
    {\def\hyper@linkstart##1##2{}%
     \let\hyper@linkend\@empty\citeyear[#1][#2]{#3}}}
\makeatother

\begin{document}

%%%%%%%%%%%%%%%%%%%%%%%%%%%%%%%%%%%%%%%%
% if you use custom commands in your title,
% ensure to check your title when submitting!
%%%%%%%%%%%%%%%%%%%%%%%%%%%%%%%%%%%%%%%%
   \title{Neutron star masses from electron-capture supernovae under equation-of-state uncertainties}
\titlerunning{Neutron star masses from ECSNe under EOS uncertainties}
%%%%%%%%%%%%%%%%%%%%%%%%%%%%%%%%%%%%%%%%
% Please separate each author with the \and command
%
% Use the \corrauth to provide the corresponding
% author address. It will be automatically inserted as 
% footnote in the PDF output.
%
% Please DO NOT include ORCIDs next to author names.
% Instead, please provide an active address for each coauthor:
% it will be automatically extracted by EDPS editorial system, 
% and co-authors will be be able to authenticate their ORCID.
%
% Only authenticated ORCIDs will be taken into account.
% ORCIDs included here will be removed.
%%%%%%%%%%%%%%%%%%%%%%%%%%%%%%%%%%%%%%%%
\author{
Vishal Parmar\inst{1}\corrauth{vishal.parmar@pi.infn.it}
\and Domenico Scordino\inst{2}
\and Ignazio Bombaci\inst{1,2}
}

\institute{
INFN, Sezione di Pisa, Largo B. Pontecorvo 3, I-56127 Pisa, Italy
\and Dipartimento di Fisica, Universit\`{a} di Pisa, Largo B. Pontecorvo 3, I-56127 Pisa, Italy
}
%   \date{Received September 30, 20XX}

% \abstract{}{}{}{}{}
% 5 {} token are mandatory
 
\abstract{
Electron-capture supernovae (ECSNe) are a promising formation channel for low-mass neutron stars, but the minimum gravitational mass of the neutron stars they produce depends on both the progenitor core structure and the neutron-star matter equation of state (EOS).
In this work, we compute the electron-capture (EC) threshold gravitational mass ($M^\star_{WD}$)  
of cold white-dwarf-like O--Ne--Mg cores with representative compositions and map the corresponding baryon number onto cold neutron-star configurations constructed from a Bayesian ensemble of unified crust--core EOSs. 
Although the EC threshold density is sensitive to the concentrations of the O--Ne--Mg mixture, the threshold baryon number of the O--Ne--Mg core varies only weakly, producing a narrow remnant-mass window.  In the baseline case with no baryonic mass loss during the transition from the EC threshold mass O--Ne--Mg core to the remnant neutron star, standard ECSNe yield remnant neutron stars with gravitational masses of $1.24$--$1.265\, M_\odot$, with only a small EOS-induced spread. Small baryonic mass losses of $0.01$--$0.02\, M_\odot$ shift this range modestly downward, but the $1.174\, M_\odot$ companion of PSR J0453+1559 would require an extreme mass loss close to $0.10\, M_\odot$, which is not favored by current ECSN simulations. We further find that the residual EOS dependence of the remnant mass is controlled mainly by the pressure around nuclear saturation density, while the corresponding tidal deformability remains sensitive to the remnant radius and compactness. Thus, low-mass double neutron star systems can in principle connect ECSN-like formation channels with gravitational-wave constraints on the EOS. Our results show that ECSNe naturally form low-mass neutron stars, but within a restricted mass range; the lightest observed neutron stars likely require low-mass iron-core collapse, ultra-stripped supernovae, or other nonstandard channels.
}

\maketitle
\nolinenumbers

%%%%%%%%%%%%%%%%%%%%%%%%%%%%%%%%%%%%%%%%%%%%%%%%%%%%%%%%%%%%%%
\section{\label{sec:introduction} Introduction}

Electron-capture supernovae (ECSNe) occupy the narrow but astrophysically important transition between white-dwarf formation and ordinary iron-core-collapse supernovae. They are expected to arise from stars near the lower edge of the massive-star regime that, after central carbon burning, develop strongly electron-degenerate  $^{16}$O--$^{20}$Ne--$^{24}$Mg cores (hereafter ONeMg cores) and evolve through the super-asymptotic giant branch phase \citep{Miyaji_1980, Nomoto1982, Poelarends:2007ip}. The existence of this channel has been recognized for decades, but modern stellar-evolution calculations have shown that its realization depends on a delicate interplay between core growth, mass loss, off-center burning, convective and mixing processes, URCA cooling, and weak-interaction rates \citep{Ray:1984vvk, Pumo:2009xj, Takahashi:2013ena, Jones:2013wda, Zha:2019rpq, Leung:2019phz, Limongi:2023bcg, Wang:2025bwm}. Because ECSNe occur at the boundary between degenerate-core survival and dynamical collapse, they offer a particularly sensitive probe of both late stellar evolution and the microphysics of dense interacting matter.

In the classical picture, the degenerate ONeMg core approaches a threshold gravitational mass ($M^\star_{WD}$) at which electron captures (ECs) on nuclei such as $^{24}$Mg and $^{20}$Ne reduce the electron fraction and the electron degeneracy pressure, thereby softening the core equation of state (EOS) and driving contraction toward collapse \citep{Miyaji_1980, Nomoto1982, Nabi:2007ek, Suzuki:2015iry, Suzuki:2020ozk}. Under these conditions, oxygen ignites degenerately near the center and launches an O--Ne deflagration \citep{ Nomoto_1984, Suzuki:2022vwq, Langanke:2015iaa}. However, the modern picture is subtler than this simple threshold argument. The ignition density, thermal structure, flame propagation, and weak rates can all alter the fate of the core, so the transition from ONeMg-core evolution to collapse is not set by a single universal threshold mass but by a physically narrow yet nonzero window in composition and structure space \citep{Langanke:2014rya, Langanke:2020gbk, Jones:2013wda, Takahashi:2018lgj, Zha:2019rpq, Leung:2019phz, Schwab:2020wte, Holas:2026rff}.

A second major development in ECSN theory was the realization that EC-triggered evolution does not necessarily end in neutron-star formation. Multidimensional studies of O--Ne  deflagrations have suggested that the system can either collapse into a neutron star or instead undergo a thermonuclear event that leaves a bound remnant, such as an O--Ne--Fe white dwarf, or more substantially disrupts the star \citep{Jones:2016asr, Jones:2018ule, Schwab:2020wte, Holas:2026rff}. The outcome depends sensitively on various factors, such as  the ignition density, conductive flame speed, and the details of flame development in degenerate O--Ne matter \citep{Jones:2016asr, Jones:2018ule}. 
ECSN theory is therefore best viewed as a bifurcation problem between collapse and thermonuclear disruption, rather than as a single deterministic channel. This point is central to any attempt to identify very low-mass neutron stars as unique products of ECSNe. An EC-triggered ONeMg core does not necessarily collapse to a neutron star. 

For those progenitors that do collapse, hydrodynamic simulations have identified several characteristic features of ECSNe. Because ONeMg cores are surrounded by steep density gradients, shock revival can occur relatively early, leading to weak explosions, low ejecta masses, low fallback, distinct neutrino signals, and characteristic nucleosynthesis \citep{Kitaura:2005bt, Wanajo:2008bw, Hudepohl:2009tyy, Wanajo:2010ig, Wanajo:2013je, Wanajo:2013jwa, zha_2022}. These features have made ECSNe attractive candidates for explaining low-kick neutron stars and a possible low-mass subpopulation in the neutron-star birth-mass distribution \citep{Podsiadlowski:2003py, Schwab:2010jm, Gessner:2018ekd, Popov:2025hva}. They have also been discussed extensively in the context of binary evolution, double neutron-star (DNS) formation, and compact-object population synthesis \citep{Poelarends:2017dua, Tauris:2015xra, Tauris:2015wla, Giacobbo:2018etu, Beniamini:2015uta, Chruslinska:2016ihp, Guo:2024xjo, Taani:2025xny}.

At the same time, one of the strongest traditional arguments for ECSNe, namely that very low neutron-star masses would provide a nearly unique observational signature of this channel, has weakened considerably. Recent neutrino-radiation-hydrodynamics simulations have shown that low-mass iron-core progenitors can also produce very light neutron stars with explosion properties that may overlap with those expected from ECSNe \citep{muller_2025, suwa_2018, Radice:2017ykv, Muller:2017gyq, Stockinger:2020hse, Janka:2025tvf}.  Moreover, stripped-envelope and ultra-stripped channels can also yield weak explosions and small natal kicks, further blurring the observational distinction between ECSNe and other low-mass core-collapse events \citep{Tauris:2015xra, Tauris:2015wla, Moriya:2016hyg, Chanlaridis:2022egz}. The low-mass end of the neutron-star mass distribution can therefore no longer be interpreted as an automatic fingerprint of EC collapse. Instead, one must quantify how low of a remnant mass the ECSN channel can actually reach once both progenitor-side and remnant-side uncertainties are treated consistently.

Observationally, the robust identification of ECSN candidates remains difficult. A major breakthrough came with SN\,2018zd, which currently provides the strongest observational case for an ECSN based on the inferred progenitor, circumstellar environment, low explosion energy, and nucleosynthetic signatures \citep{Hiramatsu:2020obu, Zhang:2020hqm}.    Synthetic light-curve and spectral modeling studies have examined the extent to which ECSNe can be distinguished observationally from low-mass Fe-core-collapse supernovae and other faint transients, while also emphasizing that the relevant diagnostics are subtle and model-dependent \citep{Tominaga:2013ala, Moriya:2014uaa, Kozyreva:2021wxn, Sato:2024gbc, Rose:2024nos}. This is one reason why neutron-star remnant properties, especially precisely measured gravitational masses, are potentially valuable as an additional diagnostic of the explosion channel.

A particularly intriguing clue comes from neutron stars with accurately measured masses significantly below the canonical $1.3$--$1.4\, M_\odot$ range. Such systems are often discussed as possible ECSN descendants because the collapse of the ONeMg core initiated at the EC mass threshold, followed by a weak explosion with little fallback, naturally suggests a low remnant mass \citep{Schwab:2010jm, Tauris:2019sho}. However, the observed gravitational mass of a neutron star is not determined by the progenitor core mass alone. Even if the stellar baryonic mass at collapse were narrowly distributed, the final gravitational mass would still depend on the binding energy of the cold remnant, and hence on the neutron-star EOS.  This baryonic-to-gravitational mass mapping is EOS-dependent \citep{Bombaci_1996}, particularly in the low-mass regime ($\simeq 1$--$1.3\, M_\odot$) relevant for ECSN remnants. Although the crust contributes only a small fraction of the stellar mass, it makes a non-negligible contribution to the radius and to the low-density structure of low-mass neutron stars. A unified treatment of the neutron-star EOS, in which the crust and core are described consistently, is therefore important for calculating the stellar binding energy and the baryonic-to-gravitational mass relation without introducing additional uncertainties from an artificial crust–core matching. Correlations among the remnant mass, radius, compactness, crust properties, and binding energy can further clarify how uncertainties in dense-matter physics affect the low-mass outcome of EC collapse.

On the progenitor side, the uncertainty is not limited to the total core mass. The internal composition of the white-dwarf-like ONeMg core, particularly the relative abundances of neon and magnesium, directly affects the threshold density for ECs, the evolution of the electron fraction, the ignition conditions for oxygen burning, and hence the baryonic mass available for neutron-star formation \citep{Poelarends:2007ip, Takahashi:2013ena, Jones:2013wda, Zha:2019rpq}. In practice, the ONeMg-core composition predicted by stellar-evolution calculations is not unique, since it depends on prior burning stages, mixing assumptions, and the treatment of weak reactions \citep{Leung:2019phz, Limongi:2023bcg, Wang:2025bwm}. Observationally, this composition is only weakly constrained and is inferred mainly through progenitor and explosion modeling rather than being measured directly. The uncertainty in the Ne/Mg mixture therefore remains largely simulation-driven, which strongly motivates exploring multiple representative core configurations.

In this work, we revisit neutron-star formation through the ECSN channel by explicitly propagating uncertainties at both stages. The central aim is to determine how tightly the ECSN remnant-mass scale can be constrained once uncertainties in the collapsing ONeMg core and the neutron-star EOS are treated simultaneously. On the progenitor side, we modeled degenerate ONeMg cores with four representative and physically motivated compositions that are commonly adopted in ECSN simulations \citep{Jones:2018ule, schwab_2015, Wang:2025bwm}: a $100\%$ $^{20}$Ne composition, a $90\%$ $^{20}$Ne + $10\%$ $^{24}$Mg mixture, a $50\%$ $^{20}$Ne + $50\%$ $^{24}$Mg mixture, and a $65\%$ $^{16}$O+$35\%$ $^{20}$Ne mixture. For each composition, we determined the onset of ECs self-consistently and obtained the corresponding threshold baryon number of the collapsing core. On the remnant side, we employed a Bayesian ensemble of unified neutron-star EOSs, which enabled us to propagate EOS uncertainties into the mass–radius relation, crust properties, compactness, conversion energy, and ultimately the baryonic-to-gravitational mass relation. The progenitor composition sets the EC threshold and the baryon number available at collapse, whereas the neutron-star EOS determines the binding energy and hence the conversion of this baryon number into the observable gravitational mass. This joint treatment allows us to establish whether the strong composition dependence of the EC threshold density leads to a comparable uncertainty in the final neutron-star mass, or whether the ECSN channel instead predicts a much narrower remnant-mass window.

We do not aim to argue that every low-mass neutron star must originate from an ECSN. Such a conclusion would be too strong in view of recent low-mass iron-core-collapse simulations and binary-evolution studies \citep{muller_2025, suwa_2018}. Rather, we ask: once realistic uncertainties in the ONeMg-core composition and in the unified neutron-star EOS are both included, what minimum remnant gravitational mass can the ECSN channel produce, and what is the corresponding plausible mass range? Answering these questions is essential for assessing whether the lowest observed neutron-star masses can be accommodated within the ECSN channel or whether alternative formation pathways are required.

The paper is organized as follows. In Sect.~\ref{sec:formalism} we describe the white-dwarf and neutron-star EOSs, the treatment of EC thresholds in degenerate ONeMg cores, and the construction of unified neutron-star models. In Sect.~\ref{sec:results} we present the resulting collapse thresholds, the correlations with crust and EOS properties, and the mapping from white-dwarf-like progenitors to neutron-star gravitational masses. Finally, in Sect.~\ref{sec:summary} we summarize our conclusions and discuss their implications for the origin of the lowest-mass neutron stars. 

\section{Formulation}
\label{sec:formalism}

We determined the onset of EC instability in white-dwarf-like ONeMg cores and combined it with a previously obtained Bayesian ensemble of unified neutron-star EOSs to infer the gravitational masses of ECSN remnants. Since the detailed construction of the neutron-star EOSs and their Bayesian inference has already been presented in our earlier work \cite{parmar_crust_baysian}, here we summarize only the ingredients needed for the present calculation and refer the reader to those papers for further details.

\subsection{White-dwarf-like ONeMg core}
\label{subsec:wd}

As discussed in Sect.~\ref{sec:introduction}, the progenitors of ECSNe evolve to form strongly degenerate, white-dwarf-like ONeMg cores. We considered unmagnetized stellar matter at zero temperature, appropriate for white dwarfs whose temperature lies below the crystallization temperature \citep{Potekhin_2009}. The atoms were assumed to be fully ionized \citep{Haensel_2007_book}, and the ions were taken to form a regular crystalline lattice. We modeled the white-dwarf-like core as a two-component plasma of nuclei 
$^{A}_{Z}X$ and $^{A'}_{Z'}X'$ embedded in a degenerate electron gas.

Under these assumptions, the total energy density can be written as
\begin{equation}
\label{eq:energy_density}
%\varepsilon = n_X\, M_N(A,Z)\, c^2 + \varepsilon_e + \varepsilon_L, %one-component system
\varepsilon = n_X\, M_N(A,Z)\, c^2 + n_{X'}\, M_N(A',Z')\, c^2 + \varepsilon_e + \varepsilon_L,
\end{equation} 
where $n_X$ and $n_{X'}$ are the ion number densities of the two species, related to the electron number density by charge neutrality, $n_e = Z\,n_X +  Z'\,n_{X'} $, $M_N(A,Z)$ and $M_N(A',Z')$ are the corresponding nuclear masses, $\varepsilon_e$ is the energy density of the relativistic electron gas, and $\varepsilon_L$ is the Coulomb lattice contribution. While the electron term is given by the standard zero-temperature expression \citep{Shapiro-Teukolsky_1983}, the Coulomb lattice energy, considering point-like ions, is written as \citep{Chamel_2015, Lunney_2003}
\begin{equation}
\label{eq:col_energy}
\varepsilon_L = -C_M\,e^2 n_e^{4/3} f(Z,Z'),
\end{equation}
where $C_M$ is the lattice (Madelung) constant, $n_e$ is the electron number density, and $f(Z,Z')$ is a dimensionless function that depends on the ionic charges and the crystal structure. For a binary lattice composed of ions $^{A}_{Z}X$ and $^{A'}_{Z'}X'$, following \citet{Jog_1982}, this function can be written as
\begin{equation}
f(Z,Z')=
\frac{\alpha Z^2+\beta Z'^2+(1-\alpha-\beta)ZZ'}{\bar{Z}^{4/3}},
\end{equation}
where $\bar{Z}$ is the mean charge number of the lattice, and the coefficients $C_M$, $\alpha$, and $\beta$ depend on the lattice geometry. For the body-centered cubic (bcc) lattice,
\begin{equation}
C_M=1.444231,\qquad
\alpha=\beta=0.389821\qquad
\end{equation}
In the one-component limit, $Z'=Z$, this expression reduces to $f(Z)=Z^{2/3}$
recovering the standard result for a one-species bcc lattice. The ion mass affects the energy density but not the pressure. Additional exchange and polarization corrections to the electron gas are neglected here, since they are known to contribute only a few percent of the Coulomb term and have little impact on the present calculation \citep{Haensel_2007_book, Potekhin_2009, Chamel_2015}. The energy and the pressure then can be calculated using the standard thermodynamical relations.

\subsection{Electron-capture instability}
\label{subsec:ec}

At sufficiently high density, EC by nuclei becomes energetically favorable. 
For a two-species ionic mixture, the relevant reactions can be written as
\begin{equation}
{}^{A}_{Z}X + e^- \rightarrow {}^{A}_{Z-1}Y + \nu_e,
\qquad
{}^{A'}_{Z'}X' + e^- \rightarrow {}^{A'}_{Z'-1}Y' + \nu_e.
\end{equation}
 {Inside a hydrostatic star, the pressure varies continuously with depth \citep{Eddington_1926} and must remain continuous across a reaction layer. The EC process therefore occurs at a fixed transition pressure, $P^\ast$, although the change in composition can be accompanied by a discontinuity in the baryon density.}

At zero temperature, the Gibbs free energy per baryon ($g$), which is equivalent to the baryon chemical potential, is the relevant thermodynamic quantity for comparing different phases at a fixed pressure. The thermodynamically stable phase is the one with the lower value of
\begin{equation}
g = \frac{\varepsilon+P}{n},
\end{equation}
where $n = A n_{X} + A’ n_{X’}$ is the total baryon number density.

For a binary mixture, the EC threshold pressure, $P^\ast$, is obtained by requiring that the Gibbs free energy per baryon of the parent composition be equal to that of the daughter composition at the same pressure,
\begin{equation}
\label{eq:gibbs}
g(A,Z;A',Z';P^\ast)=g(A,Z-\Delta Z;A',Z'-\Delta Z';P^\ast),
\end{equation}
where $\Delta Z$ and $\Delta Z'$ denote the changes in the proton numbers of the two nuclear species. For the first EC event, the relevant cases are $(\Delta Z,\Delta Z')=(1,0)$ or $(0,1)$, depending on which species becomes unstable first.

In the present work, this formalism is applied to four representative white-dwarf-like core compositions: 
a $100\%\, ^{20}$Ne composition, a $90\%\, ^{20}$Ne + $10\%\, ^{24}$Mg mixture, a $50\%\, ^{20}$Ne + $50\%\, ^{24}$Mg mixture, and 
a $65\%\, ^{16}$O + $35\%\, ^{20}$Ne mixture. These choices are intended to probe the sensitivity of the EC threshold mass $M^*_{WD}$ 
to the uncertain composition of the ONeMg core.

It is relevant to mention that the equilibrium threshold obtained here is not identical to the physical collapse condition of a realistic ECSN progenitor. In an evolving ONeMg core, ECs on the $^{24}$Mg and $^{20}$Ne chains reduce the electron fraction $Y_e$ and hence the mass that can be supported by electron degeneracy, driving further contraction and increasing the electron chemical potential. The evolution is therefore controlled by the coupled non-equilibrium evolution of density, temperature, and composition, rather than by a single equilibrium transition at fixed pressure.

In particular, realistic ONeMg cores have finite temperatures of order $T\sim10^{8},\mathrm{K}$ and are not compositionally homogeneous. Stellar-evolution calculations generally produce radial profiles of temperature, density, nuclear abundances, and electron fraction, inherited from the previous carbon-burning and shell-burning phases. During the final contraction, ECs, URCA cooling, and convective or semi-convective transport further modify these profiles, while oxygen ignition and the subsequent O–Ne deflagration can generate additional spatial variations in the thermodynamic and $Y_e$ structure that are not captured by the homogeneous model adopted here \citep{Takahashi:2013ena,schwab_2015,Zha:2019rpq,Jones:2016asr}. Coulomb and screening corrections to the weak rates also affect the capture thresholds and the subsequent evolution. Consequently, the density at which a realistic progenitor enters dynamical collapse need not coincide with the Gibbs threshold $P^\ast$ defined above.

A quantitative treatment of these effects is beyond the scope of the present work, since determining the thermal profile, weak-interaction history, composition stratification, and dynamical collapse condition of an ONeMg core requires detailed stellar-evolution calculations and, after oxygen ignition, multidimensional hydrodynamical simulations. We therefore restricted the analysis to homogeneous, zero-temperature Gibbs constructions, which provide well-defined reference threshold baryon numbers for studying how representative composition choices and neutron-star EOS uncertainties propagate into the final remnant mass.

\subsection{Unified neutron-star equation of state}
\label{subsec:nseos}

To determine the structure of the remnant neutron star and the relationship between its baryonic and gravitational masses, we employed the unified EOS framework developed in our previous work, \cite{parmar_crust_baysian}. In this approach, the crust and liquid core are constructed consistently within a single microscopic--macroscopic framework, which is particularly important for low-mass neutron stars, whose global properties are more sensitive to the crust than those of canonical-mass stars. In particular, the crust--core construction affects the radius and crust thickness and, through the resulting change in the stellar structure, also influences the compactness, binding energy, and baryonic-to-gravitational mass mapping.

The outer crust is described within the standard Baym--Pethick--Sutherland formalism \citep{BPS_1971, Haensel_2008}, where matter is modeled as a bcc lattice of nuclei embedded in a degenerate electron gas, and the equilibrium composition is obtained by minimizing the Gibbs free energy of a Wigner--Seitz (WS) cell at fixed pressure. Above neutron drip, the inner crust is treated within the compressible liquid-drop model (CLDM), following the same formalism adopted in our previous work \citep{Carreau_2019, Carreau_2020, Dinh_2021, Parmar_1}. In this region, the WS-cell energy includes bulk, surface, curvature, Coulomb, and electron contributions, with neutron-rich clusters immersed in a gas of dripped neutrons. The surface and curvature terms are parametrized as in our earlier study, with the corresponding CLDM parameters calibrated to nuclear mass data. For the bulk nuclear-matter contribution entering both the inner crust and the uniform core, we used relativistic mean-field (RMF) theory, according to which nucleons interact through the exchange of the scalar ($\sigma$), isoscalar-vector ($\omega_\mu$), and isovector-vector ($\vec{\rho}_\mu$) mesons \citep{Serot_1986, Boguta_1977, MULLER_1996, BODMER1991703, Horowitz_2001, Todd_2005, Shen_2011, Dutra_2012, Chen_2014, Vishal_2021}. The Lagrangian contains the standard nonlinear $\sigma$ self-interactions, the $\omega$ self-interaction, and the mixed isoscalar--isovector coupling, which together control the effective nucleon mass, the saturation and incompressibility properties of symmetric nuclear matter, the high-density stiffness of the EOS, and the density dependence of the symmetry energy. For a given RMF parameter set, the mean-field equations are solved to obtain the thermodynamics of homogeneous charge-neutral matter in beta equilibrium. In the liquid core, the matter consists of neutrons, protons, electrons, and muons. The crust--core transition is then determined from the crust side by the density at which the WS-cell energy density becomes equal to that of uniform neutron–proton–electron–muon (npe$\mu$) matter. Since the same RMF bulk EOS is used in both the inhomogeneous crust and the homogeneous core, the resulting neutron-star EOS is fully unified and thermodynamically consistent. For more detail, see \cite{parmar_crust_baysian}.

\subsection{Bayesian framework and posterior EOSs}
\label{subsec:bayes}

The Bayesian inference of the RMF model parameters was carried out in our earlier work, and we summarize here only its essential ingredients. In the Bayesian framework, the posterior distribution of the model parameters $\boldsymbol{\theta}$ is given by
\begin{equation}
P(\boldsymbol{\theta}\,|\,D,H)
= \frac{L(D\,|\,\boldsymbol{\theta},H)\,P(\boldsymbol{\theta}\,|\,H)}
       {P(D\,|\,H)} ,
\end{equation}
where $P(\boldsymbol{\theta}\,|\,H)$ denotes the prior distribution, $L(D\,|\,\boldsymbol{\theta},H)$ is the likelihood encoding the agreement between the model and the data, and $P(D\,|\,H)$ is the Bayesian evidence. The parameter vector $\boldsymbol{\theta}$ corresponds to the RMF coupling constants, including meson--nucleon couplings, nonlinear self-interaction terms, and isoscalar--isovector mixing parameters. We adopted broad, uniform priors that span the range of physically reasonable RMF parametrizations \citep{Dutra_2014}.

The likelihood combines a diverse set of nuclear, theoretical, and astrophysical constraints to ensure coverage over sub-saturation to supranuclear densities. At nuclear densities, we imposed constraints from empirical saturation properties and giant monopole resonance data \citep{Todd_2005}, as well as symmetry-energy information extracted from nuclear masses, electric dipole polarizability, and heavy-ion collisions \citep{Brown_2013, Tsang_2009, Danielewicz_2002}. At low densities, we included chiral effective field theory calculations for symmetric and neutron matter \citep{Drischler_2016}. At higher densities, we incorporated astrophysical constraints from  Neutron Star Interior Composition Explorer (NICER) mass--radius measurements of PSR J0030+0451 and PSR J0740+6620 \citep{Riley_2019, Miller_2019, Riley_2021, Miller2021}, as well as tidal deformability constraints from GW170817 \citep{Abbott2017}. In addition, we required that all candidate EOSs support a maximum neutron-star mass exceeding $2 M_\odot$.

The outcome of this analysis is a statistically consistent ensemble of posterior RMF parameter sets, each defining a unified EOS that consistently describes the outer crust, inner crust, and liquid core. We directly employed this posterior ensemble without repeating the inference procedure, and refer the reader to our earlier study for full details of the priors, likelihood construction, and implementation.

\subsection{Mapping the electron-capture threshold mass to the neutron star remnant mass}
\label{sec:matching}

To determine the gravitational mass of the neutron star formed in an ECSN, we followed the standard approach based on (approximate) conservation of stellar baryon number during collapse. 
The progenitor configuration was identified with the white-dwarf-like core at the EC threshold mass $M^*_{WD}$,  corresponding to a central density $\rho^*$ for the onset of ECs. 
For this configuration, the total baryon number is computed as
\begin{equation}
    N_B = \int_0^{R} n(r)\, dV \, ,
\end{equation}
where $n(r)$ is the baryon number density profile of the star and 
\begin{equation}
    dV = \left(1 - \frac{2 G m(r)}{c^2 r} \right)^{-1/2} 4\pi r^2 dr
\end{equation}
is the proper volume element and $m(r)$ is the gravitational mass enclosed within a sphere of radial coordinate $r$. The corresponding baryonic mass of the star is then defined as
\begin{equation}
    M_B = m_n N_B \, ,
\end{equation}
with $m_n$ the neutron mass. 
% This quantity represents the total rest mass of the baryons that constitute the star.  

The neutron-star remnant is obtained by constructing equilibrium configurations from the neutron-star EOS and solving the Tolman--Oppenheimer--Volkoff equations. 
The final configuration is selected by imposing baryon-number matching between the progenitor ONeMg core and the remnant neutron star, namely $N_{B,NS} = N_{B,WD}$. 
More generally, assuming some baryonic mass loss $\Delta M_B$ during the core collapse and neutron star formation we have 
\begin{equation}
N_{B,\mathrm{NS}} = N_{B,\mathrm{WD}} - \frac{\Delta M_B}{m_n}.
\end{equation}
Equivalently, the neutron star remnant baryonic mass can be written as
\begin{equation}
M_{B,NS} = M_{B,WD} - \Delta M_B.
\end{equation}
This procedure establishes a one-to-one mapping between the progenitor white-dwarf-like core and the neutron-star remnant, allowing us to determine the corresponding gravitational mass $M_{\rm NS}$. In this scenario, the stellar conversion energy ($E^{conv}$; \citealt{Bombaci-Datta_2000}) can be generalized approximately as 

%%%%%%%%%
\begin{equation}
   E^{conv} = M_G(WD, N_B^*)-M_G(NS, N_B^*- \Delta N_B)- \Delta M_B,
\label{E^conv}    
\end{equation} 
%%%%%%%%%% 
where $G$ in the subscript corresponds to the gravitational mass. 
This energy is released predominantly in the form of neutrinos during the core collapse and the subsequent de-leptonization of the newly formed neutron star \citep{Prakash_Phys_Rep_1997}. 

In realistic ECSNe, the baryonic mass loss is expected to be small because these explosions are intrinsically weak. Simulations of collapsing ONeMg cores generally yield explosion energies of order $\sim 10^{50}\,$erg, substantially lower than in ordinary iron core-collapse supernovae ($\sim 10^{51}\,$erg)\footnote{Note that the quoted values refer to the explosion kinetic energy of the ejecta, which refers to the bulk mechanical energy carried by the expanding supernova material and not to the total binding energy of the remnant neutron star.}
\citep{Kitaura:2005bt,Stockinger:2020hse,zha_2022,Wang:2025bwm,Janka:2025tvf}. This mainly reflects the steep density gradient at the edge of the ONeMg core, which causes the accretion ram pressure to decline rapidly after bounce, thereby enabling shock revival while leaving only a small amount of matter available to be unbound \cite{Kitaura:2005bt,Wanajo:2008bw,zha_2022}. Accordingly, modern multidimensional studies indicate that the directly unbound mass is typically only a few times $10^{-2}\, M_\odot$ or less \citep{Kitaura:2005bt,Wanajo:2008bw,zha_2022,Wang:2025bwm}. Motivated by these results, we took $\Delta M_B = 0$ as our baseline case, which corresponds to exact baryon-number conservation, and then examined finite values of $\Delta M_B$ separately. In particular, we considered a conservative range $\Delta M_B \sim 0.01$--$0.02\, M_\odot$, while also exploring an extreme case $\Delta M_B \sim 0.1\, M_\odot$ to assess its impact.

\section{Results and discussion}
\label{sec:results}

\subsection{Electron-capture thresholds and remnant-mass distributions}
\label{subsec:threshold_mass}

\begin{figure}
    \centering
    \includegraphics[width=1\linewidth]{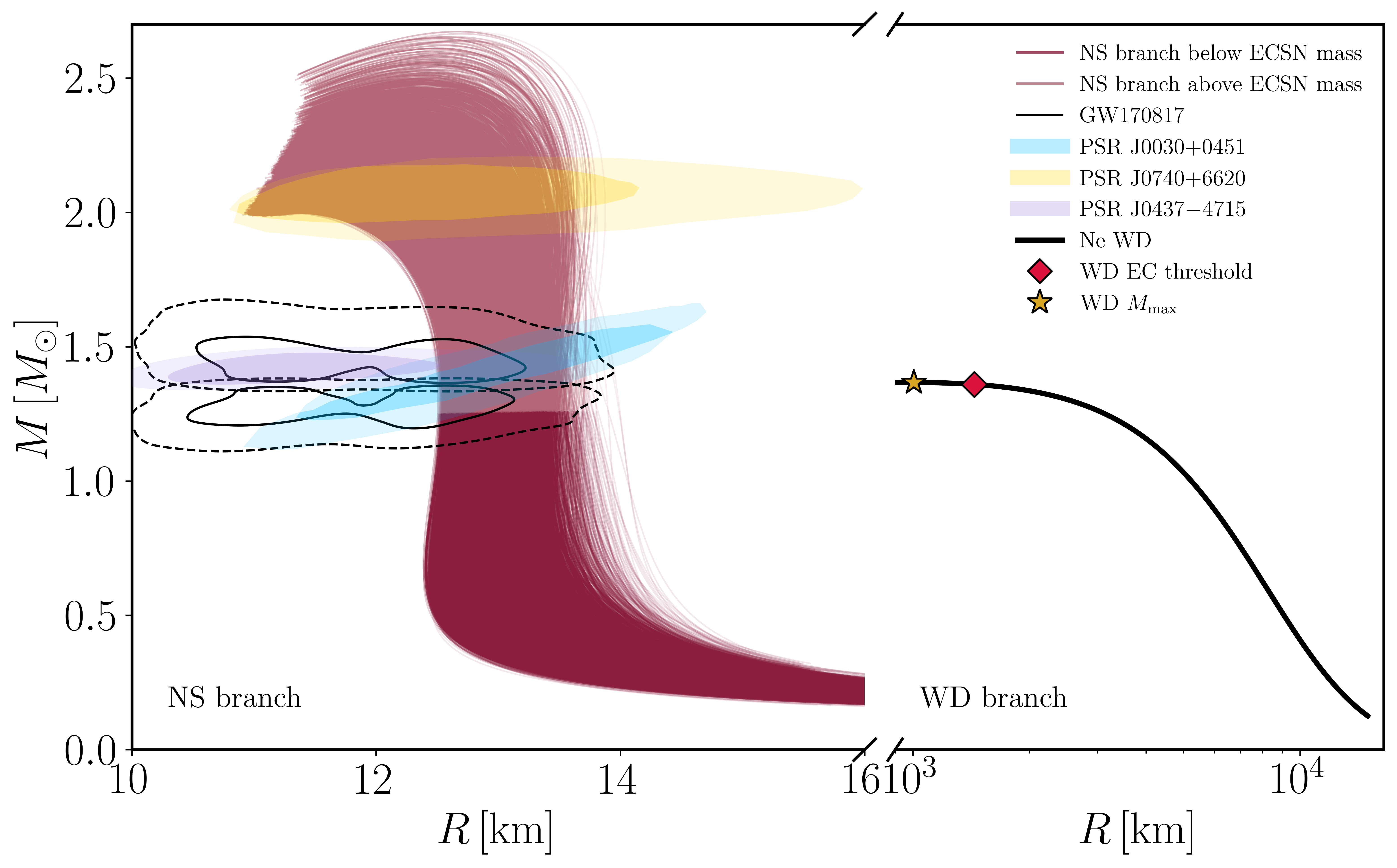}
    \caption{Gravitational mass vs. radius relations for the $^{20}$Ne white-dwarf progenitor and the corresponding neutron-star configurations from the ECSN mapping. The left and right panels show the neutron-star and white-dwarf branches, respectively. The red diamond and gold star mark the EC threshold and maximum-mass white-dwarf configurations. Observational constraints from PSR J0030+0451, PSR J0740+6620, and GW170817 are also shown.}
    \label{fig:mr}
\end{figure}

\begin{table*}
\centering
\caption{\small White-dwarf composition-dependent threshold properties.}
\label{tab:wd_ec_threshold}
\begin{tabular}{lccccc}
\hline\hline
Case &
$\rho^*$ &
$M_{\rm WD}^{*}$ &
$R_{\rm WD}^{*}$ &
$M_{\rm Ch}$ &
$N_B^{\rm *}$ \\
&
$(\mathrm{g\,cm^{-3}})$ &
$(M_\odot)$ &
$(\mathrm{km})$ &
$(M_\odot)$ &
\\
\hline
Pure $^{20}{\rm Ne}$ &
$6.809\times10^{9}$ &
$1.359634$ &
$1440.665$ &
$1.366646$ &
$1.62946\times10^{57}$ \\

$90\%\,^{20}{\rm Ne}+10\%\,^{24}{\rm Mg}$ &
$5.686\times10^{9}$ &
$1.349206$ &
$1511.432$ &
$1.368351$ &
$1.61700\times10^{57}$ \\

$50\%\,^{20}{\rm Ne}+50\%\,^{24}{\rm Mg}$ &
$3.520\times10^{9}$ &
$1.346255$ &
$1729.000$ &
$1.363866$ &
$1.61362\times10^{57}$ \\

$65\%\,^{16}{\rm O}+35\%\,^{20}{\rm Ne}$ &
$7.118\times10^{9}$ &
$1.360841 $ &
$1423.061$ &
$1.367331 $ &
$1.63085\times10^{57}$ \\
\hline\hline
\end{tabular}
\tablefoot{\small White-dwarf properties at the EC threshold for different Ne--Mg compositions. The maximum white-dwarf mass $M_{\rm Ch}$ along each sequence is also shown for reference.}
\end{table*}

\begin{figure}
    \centering
    \includegraphics[width=1\linewidth]{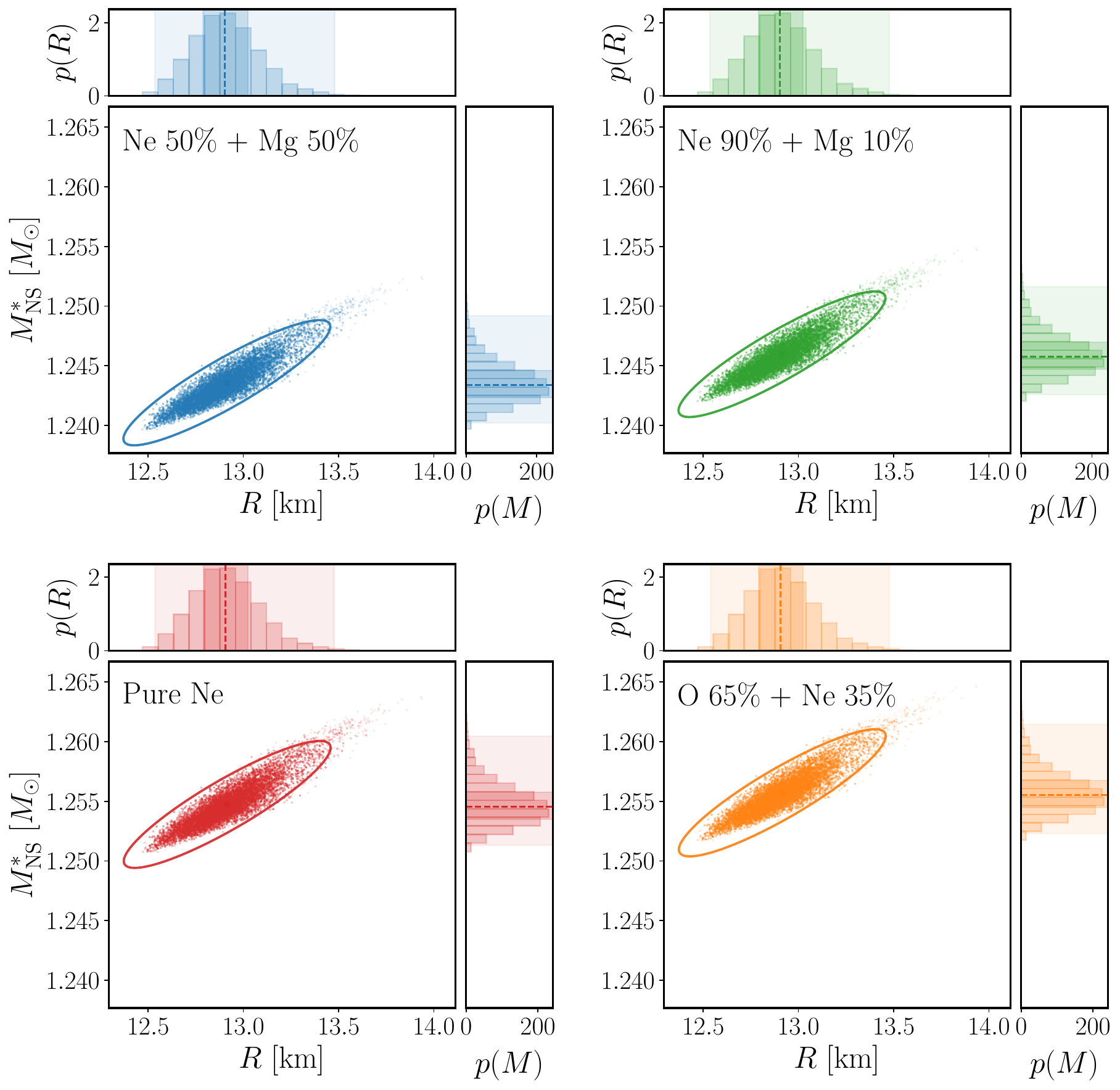}
    \caption{Predicted gravitational mass--radius distributions of neutron-star remnants formed through EC collapse for four representative progenitor compositions: $50\%$ $^{20}{\rm Ne}+50\%$ $^{24}{\rm Mg}$; $90\%$ $^{20}{\rm Ne}+10\%$ $^{24}{\rm Mg}$; pure $^{20}{\rm Ne}$; and $65\%$ $^{16}{\rm O}+35\%$ $^{20}{\rm Ne}$. The central scatter points represent the EOS-dependent remnant configurations, while the solid ellipse indicates the $99\%$ 2D Gaussian confidence region in the $R$--$M_{\rm NS}^{*}$ plane. The marginal histograms show the corresponding 1D distributions of radius and gravitational mass, with dashed lines marking the median values and shaded regions indicating the central intervals.}
    \label{fig:mr_distri}
\end{figure}

We next applied the framework described in Sect.~\ref{sec:formalism} to quantify how uncertainties in the EC threshold of ONeMg cores and in the neutron-star EOS propagate into the properties of ECSN remnants. On the progenitor side, we considered four representative cold white-dwarf-like compositions: (i) pure $^{20}$Ne; (ii)  $90\%$ $^{20}$Ne + $10\%$ $^{24}$Mg; (iii) $50\%$ $^{20}$Ne + $50\%$ $^{24}$Mg; and (iv) $65\%$ $^{16}$O + $35\%$ $^{20}$Ne. These cases represent plausible variations in the EC-active composition of degenerate ONeMg cores and lead to different threshold densities, threshold masses, and threshold baryon numbers \citep{Jones:2018ule, schwab_2015, Wang:2025bwm}. For each composition, the threshold baryon number of the white-dwarf-like core is mapped onto every member of the posterior ensemble of unified neutron-star EOSs. This selects the corresponding cold neutron-star remnant configuration and yields EOS-dependent distributions of gravitational mass, radius, compactness, and conversion energy. The resulting variation therefore, reflects both the composition dependence of the progenitor threshold and the EOS dependence of the baryonic-to-gravitational mass conversion. Since the present treatment is based on an equilibrium mapping, it is intended to isolate these two effects rather than to reproduce the full dynamical evolution of an ECSN.

Figure~\ref{fig:mr} shows the gravitational mass--radius relations for the unified neutron-star EOS ensemble adopted in this work. The EOS ensemble satisfies the nuclear-physics and astrophysical constraints imposed in~\cite{parmar_crust_baysian}, including mass--radius and tidal deformability observations. For each neutron star sequence, baryon-number conservation is used to select the neutron star configuration associated with the EC threshold of the progenitor. The red diamond on the white-dwarf-like core sequence denotes the EC threshold mass configuration for the representative 
pure $^{20}$Ne white-dwarf-like core. Once this threshold is reached, the corresponding baryon number is conserved during the collapse and is matched to the neutron-star sequence. The point on the NS sequence with the same total baryon number then gives the gravitational mass of the remnant for that EOS. The change in color tone within the band of curves representing the neutron star mass–radius relation simply marks the ECSN-remnant segment of the mass--radius curve; it should not be interpreted as a separate equilibrium branch. Applying this procedure to the full EOS ensemble yields the statistical distribution of possible ECSN remnant masses, with the spread reflecting both the uncertainty in the progenitor threshold and the neutron-star EOS.  The selected remnants occupy the low-mass neutron-star regime, where the stellar structure is especially sensitive to the low-density EOS \citep{parmar_crust_baysian, Ducoin_2011}. Using a fully unified crust--core construction avoids additional systematic shifts that can arise when an independent crust EOS is attached to a different core EOS, and therefore gives a more consistent conversion from the EC threshold baryon number to the final neutron star  gravitational mass.

Table~\ref{tab:wd_ec_threshold} summarizes the white-dwarf-like core properties at the EC threshold for the four compositions considered in our work. 
%%%pure  $^{20}$Ne, $90\%\, ^{20}$Ne + $10\%\, ^{24}$Mg, $50\%\, ^{20}$Ne + $50\%\, ^{24}$Mg  and $65\%\, ^{16}$O + $35\%\, ^{20}$Ne.   
The threshold density $\rho^\ast$ decreases with increasing $^{24}$Mg fraction, because EC on $^{24}$Mg begins at a lower density than EC on $^{20}$Ne. 
Relative to the pure-Ne case, $\rho^\ast$ is reduced by $16.5\%$ for $90\%\, ^{20}$Ne + $10\%\, ^{24}$Mg and by $48.3\%$ for 
$50\%\, ^{20}$Ne + $50\%\, ^{24}$Mg.  
The threshold mass $M_{\rm WD}^{\ast}$ and baryon number $N_B^{\ast}$ follow the same trend, but with a much weaker composition dependence: $M_{\rm WD}^{\ast}$ decreases by $0.77\%$ and $0.98\%$, while $N_B^{\ast}$ decreases by $0.77\%$ and $0.97\%$ for a 
$90\%\, ^{20}$Ne + $10\%\, ^{24}$Mg and $50\%\, ^{20}$Ne + $50\%\, ^{24}$Mg composition, respectively. 
The corresponding radius instead increases by $4.9\%$ and $20.0\%$, reflecting the fact that the instability is triggered earlier along the white-dwarf sequence, in a less compact configuration. The $65\%$ $^{16}$O + $35\%$ $^{20}$Ne composition behaves very similarly to the pure-Ne case. Its threshold density is slightly higher, by about $4.5\%$, while $M_{\rm WD}^{\ast}$ and $N_B^{\ast}$ increase only marginally, by about $0.09\%$ each. The corresponding radius is smaller by about $1.2\%$, indicating that this composition reaches the EC threshold at a slightly more compact configuration than the pure-Ne sequence. Thus, even a $10\%$ Mg admixture produces a sizable shift in the collapse density, whereas the $50\%\, ^{20}$Ne + $50\%\, ^{24}$Mg case represents a more extreme limiting configuration.

%%%%%%%%%%%%%%%%%
\begin{table}
\centering
\caption{ \small
Median and 99\% central intervals for the ECSN remnant.}
\label{tab:all_comp_99ci_loss}
\renewcommand{\arraystretch}{1.30}
\setlength{\tabcolsep}{3pt}
\resizebox{\columnwidth}{!}{%
\begin{tabular}{lcccc}
\hline\hline
Composition &
$\Delta M_B$ &
$M_{\rm NS}^{*}$ &
$R$ &
$E^{conv}$ \\
&
$[M_\odot]$ &
$[M_\odot]$ &
$[\mathrm{km}]$ &
$[10^{53}\,\mathrm{erg}]$ \\
\hline

\multirow{6}{*}{\shortstack{Pure\\Ne}}
& $0.00$ & $1.25455^{+0.00593}_{-0.00324}$ & $12.905^{+0.571}_{-0.369}$ & $1.8781^{+0.0578}_{-0.1060}$ \\
& $0.01$ & $1.24611^{+0.00586}_{-0.00319}$ & $12.904^{+0.575}_{-0.368}$ & $1.8502^{+0.0570}_{-0.1047}$ \\
& $0.02$ & $1.23766^{+0.00579}_{-0.00314}$ & $12.902^{+0.578}_{-0.367}$ & $1.8225^{+0.0562}_{-0.1035}$ \\
& $0.03$ & $1.22920^{+0.00573}_{-0.00310}$ & $12.900^{+0.581}_{-0.367}$ & $1.7950^{+0.0554}_{-0.1025}$ \\
& $0.05$ & $1.21224^{+0.00559}_{-0.00301}$ & $12.897^{+0.585}_{-0.365}$ & $1.7406^{+0.0538}_{-0.0999}$ \\
& $0.10$ & $1.16965^{+0.00530}_{-0.00282}$ & $12.887^{+0.606}_{-0.362}$ & $1.6081^{+0.0504}_{-0.0947}$ \\[2pt]
\hline

\multirow{6}{*}{\shortstack{Ne 90\%\\Mg 10\%}}
& $0.00$ & $1.24577^{+0.00585}_{-0.00319}$ & $12.904^{+0.575}_{-0.368}$ & $1.8581^{+0.0570}_{-0.1046}$ \\
& $0.01$ & $1.23732^{+0.00579}_{-0.00314}$ & $12.902^{+0.578}_{-0.367}$ & $1.8305^{+0.0561}_{-0.1035}$ \\
& $0.02$ & $1.22886^{+0.00573}_{-0.00310}$ & $12.900^{+0.582}_{-0.367}$ & $1.8029^{+0.0554}_{-0.1024}$ \\
& $0.03$ & $1.22038^{+0.00567}_{-0.00305}$ & $12.898^{+0.583}_{-0.366}$ & $1.7757^{+0.0545}_{-0.1013}$ \\
& $0.05$ & $1.20340^{+0.00551}_{-0.00297}$ & $12.895^{+0.590}_{-0.364}$ & $1.7216^{+0.0531}_{-0.0984}$ \\
& $0.10$ & $1.16076^{+0.00524}_{-0.00278}$ & $12.885^{+0.609}_{-0.361}$ & $1.5903^{+0.0496}_{-0.0936}$ \\[2pt]
\hline

\multirow{6}{*}{\shortstack{Ne 50\%\\Mg 50\%}}
& $0.00$ & $1.24338^{+0.00583}_{-0.00317}$ & $12.903^{+0.576}_{-0.368}$ & $1.8386^{+0.0567}_{-0.1043}$ \\
& $0.01$ & $1.23493^{+0.00577}_{-0.00313}$ & $12.902^{+0.580}_{-0.367}$ & $1.8109^{+0.0559}_{-0.1032}$ \\
& $0.02$ & $1.22646^{+0.00571}_{-0.00309}$ & $12.900^{+0.583}_{-0.367}$ & $1.7835^{+0.0551}_{-0.1021}$ \\
& $0.03$ & $1.21798^{+0.00565}_{-0.00304}$ & $12.898^{+0.582}_{-0.365}$ & $1.7563^{+0.0543}_{-0.1009}$ \\
& $0.05$ & $1.20100^{+0.00549}_{-0.00296}$ & $12.894^{+0.591}_{-0.363}$ & $1.7024^{+0.0529}_{-0.0980}$ \\
& $0.10$ & $1.15833^{+0.00521}_{-0.00276}$ & $12.884^{+0.610}_{-0.361}$ & $1.5713^{+0.0494}_{-0.0931}$ \\[2pt]
\hline

\multirow{6}{*}{\shortstack{O 65\%\\Ne 35\%}}
& $0.00$ & $1.25552^{+0.00594}_{-0.00324}$ & $12.905^{+0.570}_{-0.369}$ & $1.8822^{+0.0580}_{-0.1062}$ \\
& $0.01$ & $1.24708^{+0.00587}_{-0.00319}$ & $12.904^{+0.575}_{-0.368}$ & $1.8543^{+0.0571}_{-0.1048}$ \\
& $0.02$ & $1.23863^{+0.00580}_{-0.00315}$ & $12.902^{+0.578}_{-0.367}$ & $1.8266^{+0.0563}_{-0.1037}$ \\
& $0.03$ & $1.23018^{+0.00574}_{-0.00311}$ & $12.901^{+0.582}_{-0.367}$ & $1.7990^{+0.0555}_{-0.1026}$ \\
& $0.05$ & $1.21322^{+0.00560}_{-0.00301}$ & $12.897^{+0.584}_{-0.365}$ & $1.7446^{+0.0539}_{-0.1001}$ \\
& $0.10$ & $1.17064^{+0.00531}_{-0.00283}$ & $12.887^{+0.606}_{-0.362}$ & $1.6120^{+0.0505}_{-0.0948}$ \\
\hline\hline
\end{tabular}%
}
\tablefoot{\small Shown are the ECSN remnant  gravitational mass, radius, and conversion energy, shown for different progenitor compositions and assumed baryonic mass losses.}
\end{table}
%%%%%%%%%%%%%%%%%%%%%%%%%%%%%%%%%%%%%

Figure~\ref{fig:mr_distri} shows the predicted gravitational mass--radius distributions of neutron-star remnants formed through EC collapse for the four progenitor compositions, assuming no baryon loss during collapse. Each point corresponds to the remnant configuration selected from one unified neutron-star EOS by baryon-number conservation. The spread within each panel therefore reflects the uncertainty on the neutron-star EOS side. For a fixed progenitor composition, the EOS-induced variation in the remnant gravitational mass is small, on the order of $\sim 0.01,M_\odot$. The lowest remnant masses are obtained for the $50\%$ $^{20}$Ne + $50\%$ $^{24}$Mg case, reaching about $1.240,M_\odot$, while the $90\%$ $^{20}$Ne + $10\%$ $^{24}$Mg and pure $^{20}$Ne cases give minimum values of about $1.242$ $M_\odot$ and $1.25$ $M_\odot$, respectively. The $65\%$ $^{16}$O + $35\%$ $^{20}$Ne case lies very close to the pure-$^{20}$Ne result, but is shifted slightly toward higher remnant masses.  This ordering follows the threshold baryon number obtained for the corresponding 
white-dwarf configurations: increasing the $^{24}$Mg fraction lowers the EC threshold and slightly reduces the baryonic mass  
available to form the neutron star. The overall shapes of the four distributions remain very similar. This is because the threshold baryon number $N^\ast_{B,{\rm WD}}$ depends only weakly on the simplified core composition, despite the much larger variation in the threshold density. Moreover, across the neutron-star EOS posterior, the baryonic-to-gravitational mass relation exhibits only a limited dispersion in the relevant low-mass range. Consequently, the ECSN mapping primarily shifts the remnant mass distribution by a small amount, without substantially modifying its shape. The ellipses in Fig.~\ref{fig:mr_distri} summarize this behavior in the $R$--$M_{\rm NS}^{*}$ plane, while the marginal histograms show that the predicted remnant masses occupy a narrow low-mass neutron-star range.

The narrow EOS-induced spread in the remnant mass can be understood from the density range sampled by these low-mass neutron-star configurations. For each progenitor composition, the baryon-number mapping selects neutron-star configurations in the narrow low-mass regions. These low-mass configurations are significantly influenced by  the EOS around the saturation density, where in the sub-saturation region, the crustal EOS become paramount. These density regimes are already relatively well constrained in the present posterior by the combined information from empirical nuclear-matter properties, heavy-ion-collision observables, symmetry-energy constraints, and chiral effective field theory calculations at sub-saturation and near-saturation densities \citep{parmar_crust_baysian}. Their structure is therefore comparatively less sensitive to the poorly constrained high-density EOS that becomes important for the most massive neutron stars.  The resulting allowed variation in the stellar compactness at fixed baryon number is therefore limited, which leads to only a small EOS-dependent dispersion in the remnant gravitational mass. The use of unified crust–core EOSs is important in this respect as it ensures that the crust, transition region, and liquid core are described consistently for every EOS, avoiding additional systematic scatter associated with matching an independently chosen crust EOS to a different core EOS \citep{parmar_crust_baysian}.

Table~\ref{tab:all_comp_99ci_loss} summarizes the remnant properties for all progenitor compositions and assumed baryonic mass losses. The no-loss case corresponds to the direct baryon-number mapping shown in Fig.~\ref{fig:mr_distri}. We then included $\Delta M_B=0.01, 0.02\, M_\odot$ as a small-loss case motivated by hydrodynamical ECSN simulations, which typically find weak explosions and small ejecta masses. For example, recent 2D ECSN simulations obtain ejecta masses of $\sim 0.017$--$0.018\, M_\odot$, explosion energies of $\sim 1.36$--$1.48\times10^{50}$ erg, and proto-neutron star baryonic masses of $\sim1.34$--$1.357\, M_\odot$ depending on the progenitor model and weak-interaction treatment \citep{zha_2022}. Earlier ECSN simulations similarly found weak explosions with neutron-star baryonic masses close to $1.36\, M_\odot$ and little nickel production \citep{Kitaura:2005bt}. We also show the larger values $\Delta M_B=0.05$ and $0.10\, M_\odot$ as extreme limiting cases, useful for testing how much baryonic loss would be required to reach the lightest observed neutron-star masses.

For $\Delta M_B = 0$, the predicted remnant masses occupy a narrow interval. The median masses are $1.255$ $M_\odot$, $1.246$ $M_\odot$, $1.243$ $M_\odot$, and $1.256$ $M_\odot$ for pure $^{20}{\rm Ne}$, $90\%$ $^{20}{\rm Ne}+10$ $ ^{24}{\rm Mg}$, $50\%$ $^{20}{\rm Ne}+50\%$ $^{24}{\rm Mg}$, and $65\%$ $^{16}{\rm O}+35\%$ $^{20}{\rm Ne}$, respectively. The 99\% intervals remain very narrow, with the upper edge reaching only $\simeq 1.26\, M_\odot$ even for the pure-$^{20}{\rm Ne}$ and O--Ne cases. Thus, within the present equilibrium mapping, standard ECSN remnants are naturally produced in the low-mass range around $1.24$--$1.265\, M_\odot$, but not above $\simeq 1.27\, M_\odot$. Increasing the assumed baryonic loss shifts all distributions almost rigidly to lower gravitational masses: for $\Delta M_B=0.01\, M_\odot$, the medians decrease by about $0.008$--$0.009\, M_\odot$, while for $\Delta M_B=0.10\, M_\odot$ they move down to $1.158$--$1.171\, M_\odot$. The radii remain close to $12.9$ km for all cases, showing that the main effect of $\Delta M_B$ is to shift the selected point along a similar low-mass part of the neutron-star sequence. The conversion energy also decreases systematically with increasing $\Delta M_B$, from $\simeq 1.84$--$1.88\times10^{53}$ erg in the no-loss case to $\simeq 1.57$--$1.61\times10^{53}$ erg for $\Delta M_B=0.10\, M_\odot$. The $65\%\,^{16}{\rm O}+35\%\,^{20}{\rm Ne}$ composition gives values nearly identical to the pure-$^{20}{\rm Ne}$ sequence, but shifted slightly upward in mass and conversion energy because of its marginally larger threshold baryon number. The numerical limits quoted above should be interpreted in the context of the unified neutron-star EOS posterior used in this work, which is obtained from a Bayesian analysis of RMF models. Alternative EOS parametrizations or inference schemes could modify the exact lower and upper bounds. Nevertheless, our posterior satisfies current nuclear and astrophysical constraints and produces mass--radius and tidal-deformability ranges broadly consistent with other Bayesian EOS studies. For this reason, while the detailed numerical values may shift slightly with the chosen formalism, the main conclusion is expected to remain unchanged.

\begin{table}
\centering
\caption{ \small Known DNS systems below $1.29\, M_\odot$.}
\label{tab:low_mass_dns}
\renewcommand{\arraystretch}{1.10}
\setlength{\tabcolsep}{1.8pt}
\begin{tabular}{lcccc}
\hline\hline
System &
Mass [$M_\odot$] &
$P_{\rm orb}$ [d] &
$e$ &
Ref. \\
\hline

J0453+1559 &
$1.174 \pm 0.004$ &
$4.07$ &
$0.1125$ &
(a) \\

J1756--2251 &
$1.230 \pm 0.007$ &
$0.3196$ &
$0.1806$ &
(b) \\

J0737--3039 (B) &
$1.2489 \pm 0.0007$ &
$0.1023$ &
$0.0878$ &
(c) \\

J1913+1102 &
$1.27 \pm 0.03$ &
$0.2063$ &
$0.0895$ &
(d) \\

J1946+2052 (A) &
$1.2838 \pm 0.0021$ &
$0.0785$ &
$0.0638$ &
(e) \\

J1946+2052 (B) &
$1.2480 \pm 0.0021$ &
$0.0785$ &
$0.0638$ &
(e) \\
J0641+0448 &
$1.269^{+0.022}_{-0.016}$ &
$3.73$ &
$0.145$ &
(f) \\

\hline\hline
\end{tabular}
\tablefoot{DNS systems with a measured or constrained component mass below, or consistent with being below,
$1.29\, M_\odot$. The quoted value corresponds to the relevant low-mass component. 
$P_{\rm orb}$  is the orbital period in days and $e$ is the orbit eccentricity. 
References: }
\tablebib{ \small
(a) \citet{Martinez2015J0453};
(b) \citet{Ferdman2014J1756};
(c) \citet{Kramer2021DoublePulsar};
(d) \citet{Ferdman2020J1913};
(e) \citet{Meng2025J1946};
(f) \citet{Yang2026J0641}.
}

\end{table}

These results provide the reference scale for comparison with the observed low-mass components in DNS systems listed in Table~\ref{tab:low_mass_dns}. DNS systems are especially useful because relativistic timing can give precise component masses, and the mass of the second-born neutron star carries information about the supernova channel. Systems with low masses, modest eccentricities, and small inferred mass loss are commonly discussed as possible products of ECSN or ultra-stripped supernova formation. The comparison suggests a clear hierarchy. The masses of PSR J0737--3039B and PSR J1946+2052B, both close to $1.249\, M_\odot$, fall directly in the no-loss ECSN range predicted here. PSR J1756--2251, with a low-mass companion of $1.230\pm0.007\, M_\odot$, can be reached if a small baryonic loss of order $\Delta M_B\simeq0.02\, M_\odot$ is allowed. In contrast, the very low mass of the companion in PSR J0453+1559, $1.174\pm0.004\, M_\odot$, cannot be obtained in the no-loss or small-loss cases. Considering the full uncertainty from both the progenitor composition and the neutron-star EOS, we find that a mass as low as that of the companion in PSR J0453+1559, $M=1.174\pm0.004\, M_\odot$, cannot be obtained within the standard ECSN mapping unless an extreme baryonic mass loss, close to $\Delta M_B\simeq0.10\, M_\odot$, is assumed.  Such a large loss is not favored by current ECSN simulations, which generally predict a rather well-defined remnant mass set by the EC threshold mass of the ONeMg core and only modest ejecta.

This conclusion is consistent with recent studies suggesting that the lightest neutron stars are more naturally produced by low-mass iron-core collapse rather than by ECSNe. In particular, \cite{suwa_2018} show that low-mass CO cores in ultra-stripped systems can form sufficiently small Fe cores and produce neutron stars with gravitational masses around $1.17\, M_\odot$, compatible with PSR J0453+1559. They further argued that, because ECSN progenitors have a higher electron fraction, ECSNe are expected to produce more massive neutron stars than low-mass Fe-core collapse events. Similarly, recent 3D simulations by \cite{muller_2025} obtained a record-low gravitational mass of $1.192\, M_\odot$ from neutrino-driven collapse of low-mass iron-core progenitors, and emphasized that ECSNe typically give a gravitational mass around $1.24\, M_\odot$ with only small EOS uncertainty. Therefore, if the $1.174\, M_\odot$ object is confirmed to be a neutron star, it is more naturally interpreted as the outcome of a low-mass iron-core collapse or ultra-stripped supernova. Alternatively, additional physics, such as dark-matter-admixed progenitors or compact remnants, may be required to reduce the effective baryonic mass before collapse \citep{PARMAR2026100470}.

At the high-mass end of the low-mass DNS sample, the situation is also restrictive. The present ECSN mapping does not naturally produce remnants above $\simeq1.27\, M_\odot$, even for the pure-Ne and $65\%\,^{16}{\rm O}+35\%\,^{20}{\rm Ne}$ case with no baryon loss. Reaching masses close to $1.29\, M_\odot$ would require a significantly smaller baryonic-to-gravitational mass conversion, which effectively corresponds to a very stiff neutron-star EOS. Such EOSs are strongly constrained by GW170817, since large radii and large tidal deformabilities are disfavored. Conversely, very soft EOSs are also restricted by NICER mass--radius measurements and by the requirement of supporting $\sim2\, M_\odot$ neutron stars. Therefore, within the observationally allowed EOS ensemble used here, systems such as PSR J1946+2052B, with $M\simeq1.284,M_\odot$, are not favored as standard ECSN remnants. PSR J1913+1102, with $M=1.27\pm0.03,M_\odot$, remains only marginally compatible, because its uncertainty overlaps the upper edge of the predicted no-loss ECSN range. PSR J0641+0448, with $M=1.269^{+0.022}_{-0.016},M_\odot$, falls in the same category: its central mass is somewhat higher than the typical predicted ECSN remnant mass, but its lower uncertainty range remains compatible with the upper tail of the no-loss prediction. A robust mass significantly above this range would point instead to a different formation channel, most likely an ultra-stripped or low-mass iron-core collapse supernova rather than a standard ECSN. This conclusion is broadly consistent with the wide-binary population-synthesis study of \citet{Stevenson:2022cmh}, where ECSN-formed pulsars in wide, non-interacting binaries are associated with a low gravitational-mass range of $1.25<M_{\rm psr}/M_\odot<1.3$. Our EOS-based baryon-number mapping naturally supports this low-mass ECSN interpretation, although the upper end of that range, near $1.3,M_\odot$, lies somewhat above the typical values predicted in our standard no-loss models.

\subsection{EOS sensitivity of ECSN remnant properties}
\label{subsec:eos_sensitivity}

\begin{figure}
    \centering
    \includegraphics[width=1\linewidth]{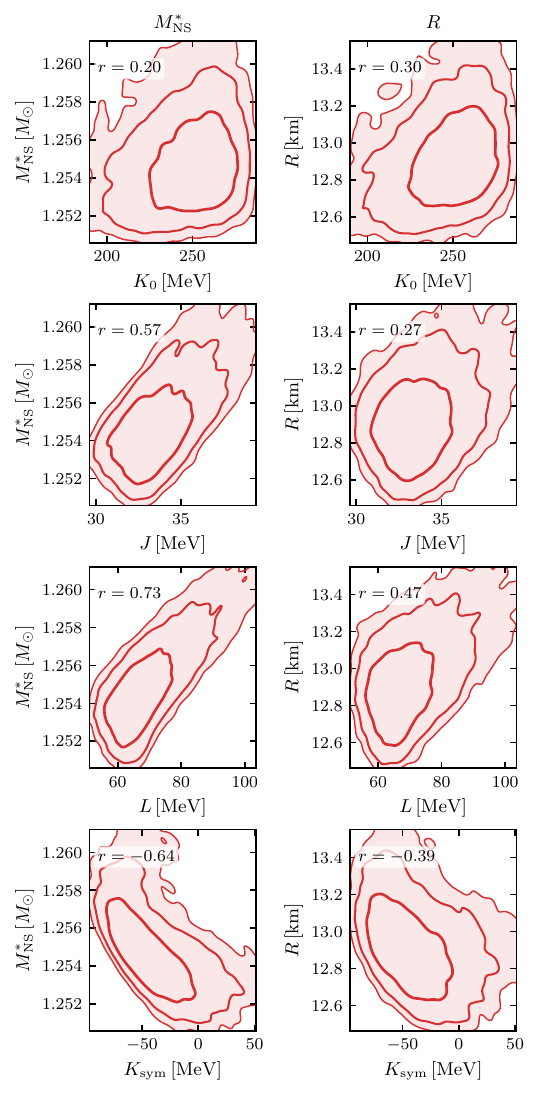}
    \caption{Correlations between the ECSN remnant properties and nuclear-matter parameters for the pure-Ne progenitor case with no baryon loss. \textit{Left column}: Gravitational remnant mass, $M_{\rm NS}^{*}$. \textit{Right}: Corresponding radius, $R$. \textit{Top to bottom}: Correlations with the incompressibility ($K_0$), symmetry energy ($J$), slope parameter ($L$), and symmetry-energy curvature ($K_{\rm sym}$). The contours enclose the 2D confidence regions of the EOS ensemble, and the Pearson correlation coefficient ($r$) is reported in each panel.
}
\label{fig:PureNe_NM_correlations}
    \label{fig:nm}
\end{figure}

\begin{figure*}
    \centering
    \includegraphics[width=1\linewidth]{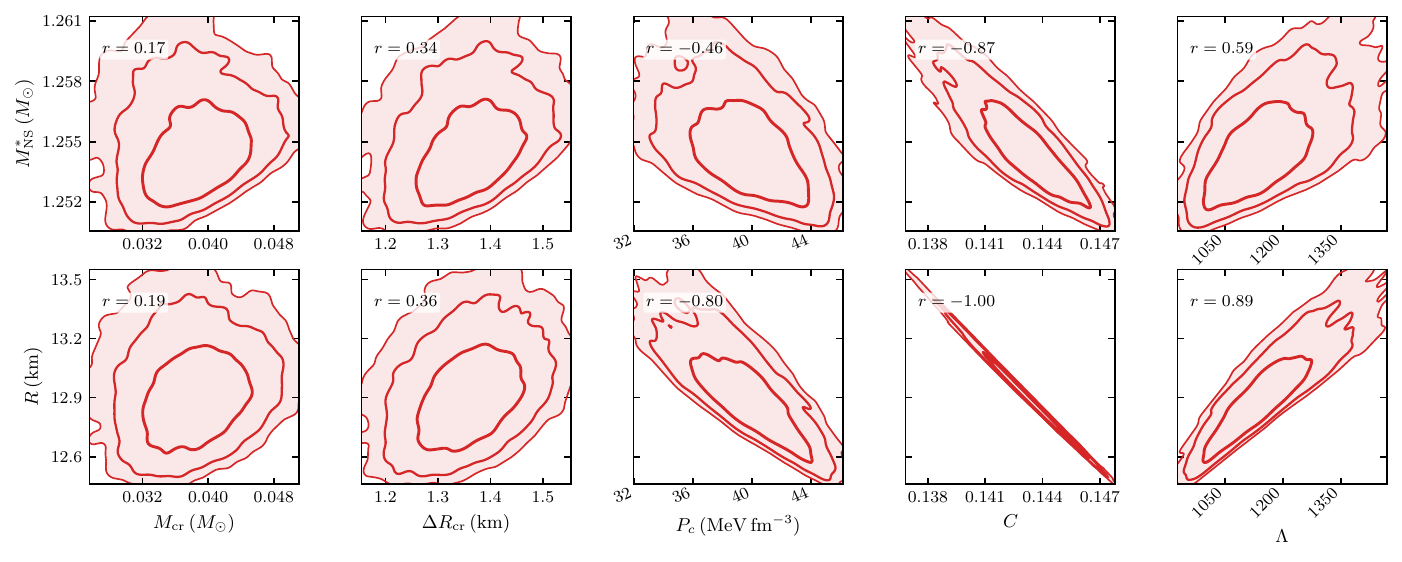}
    \caption{Correlations of the ECSN remnant gravitational mass ($M_{\rm NS}^{*}$) and radius ($R$) with selected neutron-star structural properties for the pure-Ne progenitor case without baryon loss. \textit{Top row}: Correlations with $M_{\rm NS}^{*}$. \textit{Bottom row}: Correlations with $R$. Columns from \textit{left to right}: Crust mass ($M_{\rm cr}$), crust thickness ($\Delta R_{\rm cr}$), central pressure ($P_c$), compactness ($C$), and dimensionless tidal deformability ($\Lambda$) of the selected remnant configuration. The contours show the 2D confidence regions of the EOS ensemble, and the Pearson correlation coefficient ($r$) is reported in each panel.
}
\label{fig:ECSN_MR_vs_structure}
    \label{fig:ns_corr}
\end{figure*}

Figure~\ref{fig:PureNe_NM_correlations} identifies the nuclear-matter properties that govern the residual EOS dependence of the ECSN remnant mass and radius. Because the progenitor threshold fixes the baryon number of the collapsing core, variations across the EOS ensemble affect the remnant properties only through the EOS-dependent stellar structure and binding energy. The correlations therefore indicate which nuclear-matter parameters control the conversion from the fixed progenitor baryon number to the observable gravitational mass. In Fig.~\ref{fig:PureNe_NM_correlations}, $K_0$ denotes the incompressibility of symmetric matter, $J$ the symmetry energy, $L$ its slope, and $K_{\rm sym}$ its curvature. The weak correlations with $K_0$ show that the isoscalar sector does not dominate the residual uncertainty in the remnant mass. Instead, the remnant mass is mainly controlled by the isovector sector, increasing with $J$ and especially with $L$, and decreasing with $K_{\rm sym}$. The radius follows the same qualitative trend, although with weaker correlations. Physically, a larger $L$ increases the pressure in the relevant density range, producing a larger radius and a smaller binding-energy correction at fixed baryon number; the same collapsing core is therefore mapped to a slightly larger gravitational mass. This also means that a precise mass--radius measurement of a low-mass neutron star would probe the same EOS region that enters the ECSN mass conversion. However, mass and radius alone would not uniquely establish an ECSN origin. Such an interpretation would require consistency with the predicted narrow remnant-mass range, together with independent indicators of a low-kick formation channel, such as a DNS configuration, low orbital eccentricity, small inferred mass loss, or evidence that the second supernova imparted only a weak natal kick.

We next examined how the ECSN remnant properties are connected to the internal structure of the remnant itself. In Fig.~\ref{fig:ns_corr}, we correlate the remnant gravitational mass $M_{\rm NS}^{*}$ and radius $R$ with quantities evaluated for the same selected stellar configuration: the crust mass $M_{\rm cr}$, crust thickness $\Delta R_{\rm cr}$, central pressure $P_c$, compactness $C$, and dimensionless tidal deformability $\Lambda$. The crustal quantities are obtained directly from the TOV profile of each selected remnant. The crust--core boundary is fixed by the transition pressure $P_t$ of the corresponding unified EOS; the radius $R_{\rm core}$ and enclosed gravitational mass $M_{\rm core}$ are then read from the stellar profile at $P=P_t$. We defined the crust thickness and crust mass as
$\Delta R_{\rm cr}=R-R_{\rm core}$ and $M_{\rm cr}=M_{\rm NS}^{*}-M_{\rm core}$.
Because $P_t$ is determined consistently from the same crust--core EOS construction, these quantities reflect the internal structure of the selected low-mass remnant and do not rely on an artificial crust attachment \citep{parmar_crust_baysian, Carreau_2019}.  Although fully unified EOSs are used, the crust mass $M_{\rm cr}$ and crust thickness $\Delta R_{\rm cr}$ show only weak correlations with $M_{\rm NS}^{*}$ and $R$. This is because the crust contributes only a small fraction of the stellar mass, so changes in $M_{\rm cr}$ or $\Delta R_{\rm cr}$ do not strongly affect the baryonic-to-gravitational mass conversion. The importance of the unified EOS is instead that the crust, crust--core transition, and core are treated consistently when calculating the radius and baryon number of the same low-mass remnant. This is particularly relevant for the radius, where an artificial crust--core matching could introduce systematic shifts.  The central pressure $P_c$ shows a clearer anticorrelation, especially with the radius. Remnants with larger $P_c$ are more compact and therefore have smaller radii, as expected for configurations located deeper along the same low-mass branch. The correlation with $M_{\rm NS}^{*}$ is weaker, indicating that $P_c$ mainly reflects the compactness of the selected remnant rather than producing a large shift in the gravitational mass. 

The compactness and tidal deformability show much stronger correlations with the remnant properties. The compactness is almost perfectly anticorrelated with the radius, because for these ECSN remnants the mass varies only within a narrow interval, so changes in $C=GM/(Rc^2)$ are driven mainly by changes in $R$. Thus, a smaller radius immediately corresponds to a larger compactness. A clear anticorrelation is also found between $M_{\rm NS}^{*}$ and $C$, indicating that EOSs producing slightly larger remnant masses tend to give more compact low-mass configurations. This reflects the fact that the baryonic-to-gravitational mass conversion is controlled by the neutron star’s binding energy, which is itself set by the compactness of the remnant neutron star. 
The dimensionless tidal deformability $\Lambda$ behaves in the opposite way. Since $\Lambda$ depends very strongly on the stellar radius and compactness, less compact remnants have larger tidal deformabilities, while more compact remnants have smaller $\Lambda$. This explains the strong positive correlation between $R$ and $\Lambda$, and the weaker but still visible positive correlation between $M_{\rm NS}^{*}$ and $\Lambda$. In this low-mass ECSN window, the spread in $\Lambda$ therefore carries direct information about the stiffness of the unified EOS around the densities sampled by the remnant.

\begin{figure}
    \centering
    \includegraphics[width=1\linewidth]{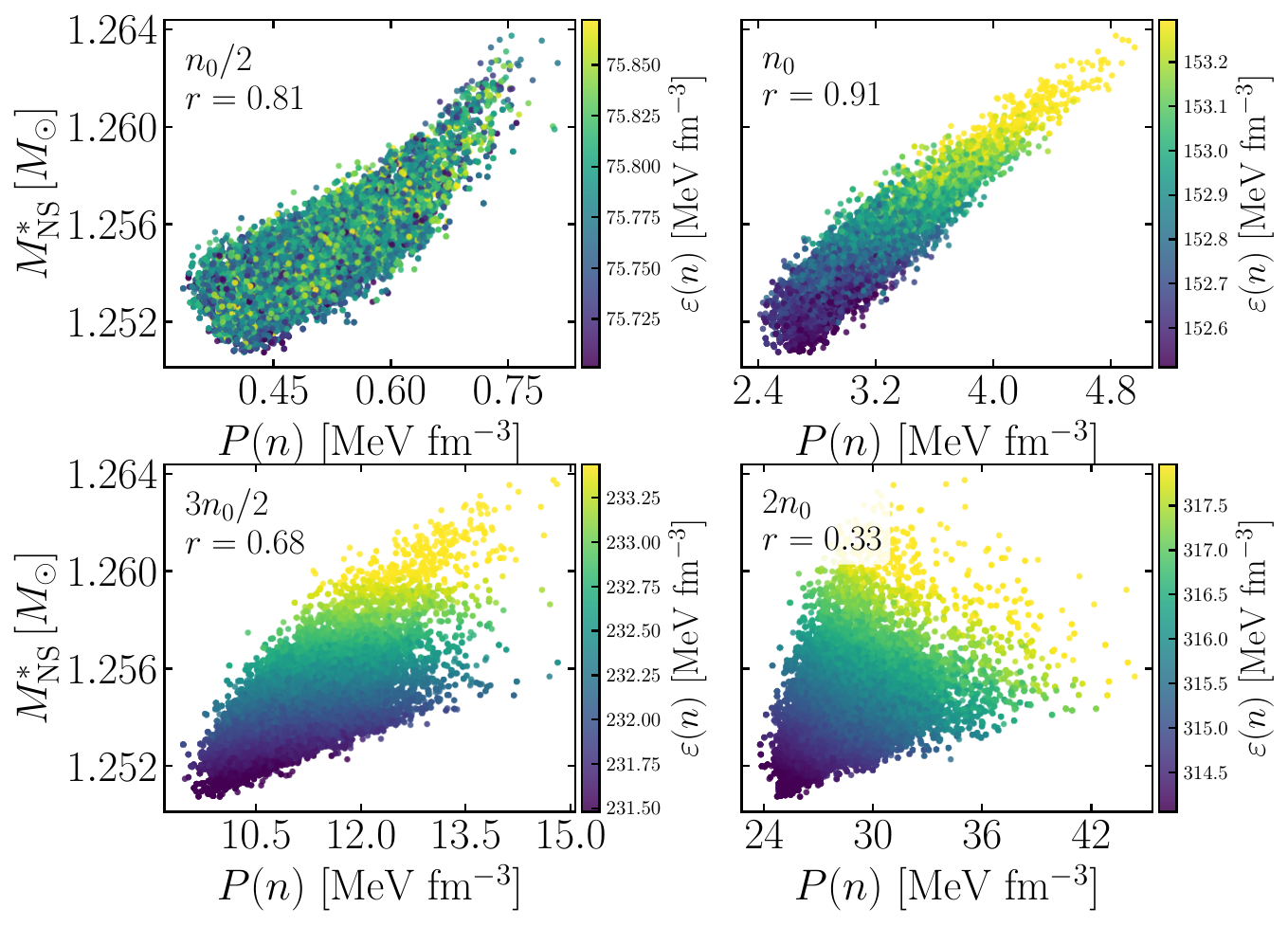}
    \caption{Correlation between the gravitational mass of the ECSN remnant, $M^\ast_{\rm NS}$, and the pressure of neutron star matter, $P(n)$, evaluated at four representative densities: $n_0/2$, $n_0$, $3n_0/2$, and $2n_0$. 
The points represent the unified-EOS ensemble, while the color scale indicates the corresponding energy density, $\epsilon(n)$, at each density. }
    \label{fig:press}
\end{figure}

 Figure~\ref{fig:press} provides a useful way to interpret the ECSN remnant mass as an EOS diagnostic. The figure correlates 
$M^\ast_{\rm NS}$  with the pressure of neutron-star matter at selected densities, with the color scale showing the corresponding energy density. The strongest correlation occurs near saturation density $n_0$, especially at $n_0$, while the correlation becomes weaker toward $2n_0$. This shows that the residual EOS uncertainty in the ECSN remnant mass is not primarily controlled by the high-density core, but by the pressure in the density interval actually sampled by these low-mass configurations. In this regime, a larger pressure around $n_0$ gives a less compact star at fixed baryon number, reduces the binding-energy correction, and therefore leads to a slightly larger gravitational mass.

This correlation can be read in two complementary ways. If a low-mass neutron star has independent formation-channel evidence pointing to an ECSN origin, for example from its membership in a DNS system, low eccentricity, and small inferred mass loss or kick, then its precisely measured gravitational mass can provide an astrophysical constraint on the near-saturation pressure of the EOS. Conversely, if nuclear experiments, chiral effective field theory, or future radius and tidal-deformability measurements narrow the allowed pressure around $n_0$, then the ECSN remnant-mass window will become correspondingly sharper. This would make it possible to decide more cleanly whether a given low-mass neutron star is compatible with standard ECSN formation or requires another channel, such as low-mass iron-core collapse or an ultra-stripped supernova.

\subsection{Tidal signatures of ECSN remnants}
\label{subsec:tidal_signatures}

\begin{figure}
    \centering
    \includegraphics[width=1    \linewidth]{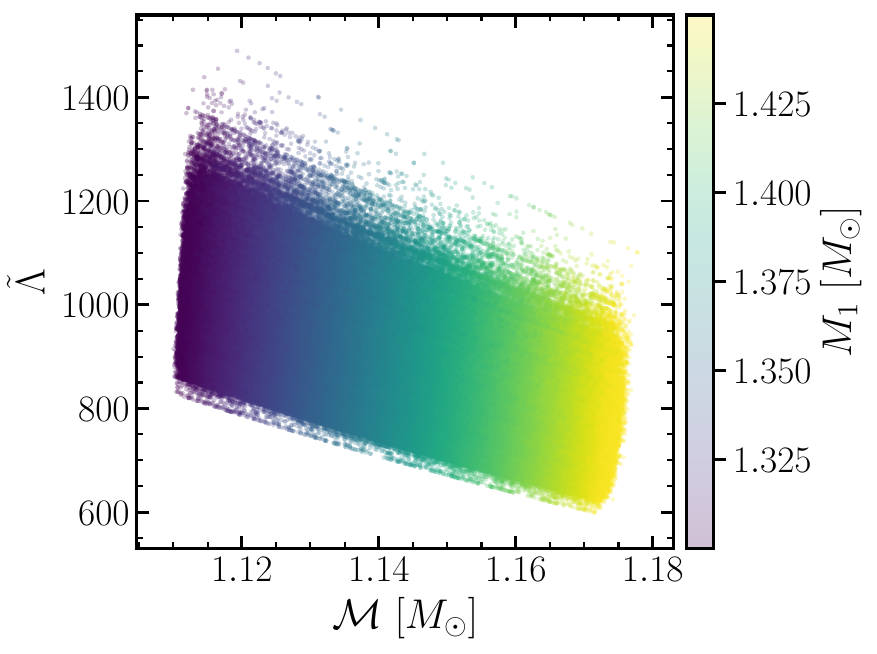}
    \caption{Effective tidal deformability ($\tilde{\Lambda}$) as a function of chirp mass ($\mathcal{M}$) for binary neutron-star systems containing an ECSN-born remnant. 
The recycled companion mass ($M_1$) is sampled over a DNS-like mass range, as indicated by the color bar; the second component is the ECSN remnant obtained from the EOS ensemble.  }
    \label{fig:tidal}
\end{figure}

The compactness and tidal-deformability correlations provide a direct connection between the ECSN-remnant calculations and binary neutron-star mergers. While the remnant mass is confined to a narrow interval by the baryon-number mapping, the radius and tidal deformability retain a larger dependence on the unified EOS. Two ECSN remnants with similar gravitational masses can therefore have appreciably different tidal deformabilities because their compactnesses differ. To connect these results with gravitational-wave observables, Fig.~\ref{fig:tidal} shows the effective tidal deformability ($\tilde{\Lambda}$) as a function of the chirp mass ($\mathcal{M}$) for binaries composed of an ECSN-born remnant and a recycled companion. The companion mass is sampled over the DNS-like range $M_1=1.30$--$1.45\, M_\odot$, while the second component is taken from the ECSN remnant distribution obtained for each unified EOS. This setup is motivated by population-synthesis studies, which have shown that ECSNe can contribute significantly to the formation of merging DNSs if they are associated with small natal kicks. In particular, \citep{Giacobbo_2019} found that the number of DNS systems generally increases for smaller ECSN kick dispersions, and that a large fraction of merging DNSs can experience at least one ECSN. At the same time, similar low-kick outcomes can also arise from related channels, such as ultra-stripped supernovae \citep{Giacobbo_2019, Tauris:2015xra}. Therefore, the low mass scale should be interpreted as evidence of compatibility with an ECSN-like channel rather than as a unique identification of the formation mechanism.

Our calculation adds the EOS-dependent counterpart to this formation picture. Rather than assuming a fixed ECSN gravitational mass, we propagated the unified-EOS dependence of the baryonic-to-gravitational mass conversion and of the remnant radius into the tidal sector. As shown in Fig.~\ref{fig:tidal}, the resulting binaries occupy a continuous band in the $\mathcal{M}$--$\tilde{\Lambda}$ plane, where $\mathcal{M}=(M_1M_2)^{3/5}/(M_1+M_2)^{1/5}$ denotes the chirp mass and $\tilde{\Lambda}=(16/13)[(M_1+12M_2)M_1^4\Lambda_1+(M_2+12M_1)M_2^4\Lambda_2]/(M_1+M_2)^5$ denotes the binary effective tidal deformability.  {For each EOS in the posterior ensemble, the same unified EOS is used to determine the tidal deformabilities of both binary components. The ECSN-remnant mass $M_2$ is obtained from the baryon-number mapping, while the recycled companion mass $M_1$ is sampled in the range $1.30$–$1.45,M_\odot$. The corresponding deformabilities, $\Lambda_1=\Lambda_{\rm EOS}(M_1)$ and $\Lambda_2=\Lambda_{\rm EOS}(M_2)$, are evaluated from the mass–tidal-deformability sequence of that same EOS before constructing $\tilde{\Lambda}$.} The spread in $\mathcal{M}$ is mainly driven by the adopted recycled-companion mass range, while the spread in $\tilde{\Lambda}$ reflects the EOS dependence of the stellar radii and compactnesses. Since tidal effects in the inspiral are encoded in $\tilde{\Lambda}$, which is strongly correlated with the stellar radius at fixed chirp mass \citep{De:2018uhw,Raithel:2018ncd}, this observable provides a direct link between the internal structure of the ECSN remnant and the gravitational-wave signal. Thus, while the chirp mass can test compatibility with a low-mass ECSN-like formation channel, $\tilde{\Lambda}$ constrains the compactness and radius predicted by the EOS.

In summary, the central result of this work is not simply that ECSNe produce low-mass neutron stars. Rather, the mass scale is set jointly by the EC threshold of the progenitor and by the near-saturation part of the unified neutron-star EOS. Standard ECSNe naturally populate a narrow range around $1.24$--$1.265\, M_\odot$ in the no-mass-loss case, with small baryonic losses shifting this range modestly downward. Masses far below this range require unrealistically large baryonic loss within the present framework, while masses significantly above it are also difficult to accommodate as standard ECSN remnants. The lowest-mass neutron stars therefore provide a useful test of both formation physics and the EOS, but only when the two are treated together.

\section{Conclusions}
\label{sec:summary}

We have studied the gravitational masses of neutron stars formed through the ECSN channel by combining white-dwarf-like ONeMg progenitor models with a Bayesian ensemble of unified neutron-star EOSs. The EC threshold was computed for four representative O--Ne--Mg compositions, and the corresponding progenitor baryon number was mapped onto cold neutron-star sequences. This allowed both 
progenitor-composition uncertainties and EOS uncertainties to be propagated into the final remnant mass.

By treating the progenitor and remnant sides within the same calculation, we find that the uncertainty in the final ECSN mass is more limited than might be expected from the variation in the EC threshold density alone. The different ONeMg compositions considered here lead to appreciable changes in the threshold density, but to comparatively small changes in the baryon number of the collapsing core. The subsequent mapping through the unified EOS ensemble introduces an additional dependence through the neutron-star binding energy, while preserving a relatively narrow range of predicted remnant masses. This provides a quantitative reference for comparing standard ECSN expectations with the masses of low-mass neutron stars in binary systems.

We find that the ECSN channel naturally produces neutron stars in a narrow low-mass range. In the baseline case with no baryonic mass loss, the predicted gravitational masses lie around $1.24$--$1.265\, M_\odot$, with only a small EOS-induced spread. Small baryonic losses of order $\Delta M_B=0.01$--$0.02\, M_\odot$ shift this range modestly downward but cannot explain the very low mass of the companion in PSR J0453+1559, $M\simeq1.174\, M_\odot$. Reaching such a mass would require an extreme baryonic loss close to $\Delta M_B\simeq0.10\, M_\odot$, which is not favored by current ECSN simulations. This supports the interpretation that the lightest neutron stars are more likely produced by low-mass iron-core collapse, ultra-stripped supernovae, or other nonstandard channels.

The remaining spread in the ECSN remnant mass is controlled mainly by the EOS pressure around nuclear saturation density, rather than by the high-density core. Thus, a precisely measured low-mass neutron star with independent evidence of an ECSN-like origin could provide an astrophysical probe of the near-saturation EOS. Conversely, improved nuclear and astrophysical constraints on this density region would sharpen the predicted ECSN mass window. In binary neutron-star systems, the remnant mass can test compatibility with an ECSN-like low-kick formation channel, while the tidal deformability carries complementary information about the remnant radius and compactness. The lowest-mass neutron stars therefore provide a joint test of progenitor evolution, baryonic mass loss, and the neutron-star EOS.

\begin{acknowledgements}
We sincerely thank the referee for a careful and thorough reading of the manuscript. The detailed comments and constructive suggestions have helped us clarify several aspects of the presentation and improve the overall quality of the work.
\end{acknowledgements}

\bibliographystyle{aa_url} % style aa.bst
\bibliography{main} % your references Yourfile.bib
\end{document}